\documentclass[aps,prb,floats,floatfix,amssymb,twocolumn,letterpaper,
groupedaddress,twoside,showpacs]{revtex4}
\usepackage{graphicx}
\usepackage{amssymb,amsmath}
\usepackage{epsfig}

\begin{document}

\title{
Low-Temperature Orientation Dependence of Step Stiffness on \{111\} Surfaces 
}


\author{T. J. Stasevich}
\email[]{tjs@glue.umd.edu}
\author{Hailu Gebremariam}
\author{T. L. Einstein}
\thanks{Corresponding author}
\email[]{einstein@umd.edu}
\homepage[]{http://www2.physics.umd.edu/~einstein/}
\affiliation{Department of Physics, University of Maryland, College Park, MD 20742-4111}
\author{M. Giesen}
\author{C. Steimer}
\author{H. Ibach}
\affiliation{Institut f\"ur Schichten und Grenzfl\"achen (ISG),
Forschungszentrum J\"ulich, D-52425 J\"ulich, Germany}


\date{\today}

\begin{abstract}
For hexagonal nets, descriptive of \{111\} fcc surfaces, we derive from combinatoric arguments a
simple, low-temperature formula 
for the orientation dependence of the surface step line tension and stiffness, as well as the
leading correction, based on the Ising model with nearest-neighbor (NN) interactions.
Our formula agrees well with experimental data for 
both Ag and Cu\{111\} surfaces, indicating that
NN-interactions alone can account for the data in these cases (in contrast to results for
Cu\{001\}). 
Experimentally significant corollaries of the low-temperature derivation show that the step line
tension cannot be extracted from the stiffness and that with plausible assumptions the
low-temperature stiffness should have 6-fold symmetry, in contrast to the 3-fold symmetry of the
crystal shape. We examine Zia's exact implicit solution in detail, using numerical methods for
general orientations and deriving many analytic results including explicit solutions in the two
high-symmetry directions.  From these exact results we rederive our simple result and explore
subtle behavior near close-packed directions.  To account for the 3-fold symmetry in a lattice
gas model, we invoke a novel orientation-dependent trio interaction and examine its
consequences.    
\end{abstract}



\maketitle

\section{Introduction}
Most of our current understanding of surface morphology is based on
the step-continuum model,\cite{JW} which treats the step itself 
as the fundamental unit controlling the evolution of a surface.
In this model, the step stiffness $\tilde{\beta}$ serves as one of the three fundamental 
parameters; it gauges the ``resistance" of the step to meandering and ultimately accounts 
for the inertia of the step in the face of driving forces.
The stiffness can be derived from the step line tension $\beta(\theta)$, the excess free energy
per length
associated with a step edge of a specified azimuthal orientation $\theta$.  From $\beta$, the
two-dimensional equilibrium crystal shape (i.e., the shape of the islands) is determined. 

The goal of this paper is to find low-temperature ($T$) formulas for   
$\beta(\theta)$ and thence $\tilde{\beta}(\theta) \equiv \beta(\theta) +\beta^{\prime
\prime}(\theta)$ as functions of the azimuthal misorientation $\theta$, assuming just
nearest-neighbor interactions in plane and an underlying \{111\} surface.
Such surfaces are 
characterized by a six-fold symmetric triangular (hexagonal) lattice, allowing all calculations
to 
be done in the first sextant alone (from $0^\circ$ to $60^\circ$).  In contrast to
$\beta(\theta)$, we shall find that at low $T$, $\tilde{\beta}(\theta)$ is insensitive, under
plausible assumptions, to the symmetry-breaking by the second substrate layer, so that plots
from $0^\circ$ to $30^\circ$ suffice.
Although an analytic solution exists for the orientation dependence of $\beta(\theta)$ on a
square 
lattice,\cite{abraham, rottman, avron} only an implicit solution to a 6$^{th}$-order equation
has been found 
for a hexagonal lattice.\cite{zia}  This makes 
comparisons to experiment rather arduous, particularly when trying 
to compare data on $\tilde{\beta}(\theta)$, which is related to $\beta(\theta)$ through a double
derivative 
with respect to $\theta$.
Fortunately, we will see that a remarkably simple formula exists for the orientation dependence
of 
$\tilde{\beta}$ at temperatures which are low compared to the characteristic energy of step
fluctuations (i.e. the kink energy or the energy per length along the step).  For noble metals,
room temperature lies in this limit, facilitating direct comparisons to experiment.    

Motivating this work is 
a recent finding\cite{dieluweit02} that the square-lattice nearest-neighbor (NN) Ising model
underestimates $\tilde{\beta}$ by a factor of $4$ away from close-packed directions on
Cu\{001\}.  Later work\cite{ZP1,ZP2,tjs001} showed that much (but not all) of this discrepancy
could be understood 
by considering the addition of next-nearest-neighbor (NNN) interactions.  For the triangular
lattice, we will see that 
such a longer-range interaction is not required to
describe the orientation dependence of $\tilde{\beta}$.       

In the following Section, 
we characterize steps on a hexagonal lattice and perform a low-temperature 
expansion of the lattice-gas partition function, assuming only NN bonds 
are relevant, and derive both $\beta(\theta)$ and $\tilde{\beta}(\theta)$. 
We obtain a remarkably simple expression for the latter in Eq.~(\ref{eq:stiffness}). Since this
low-$T$ limit is determined by geometric/configurational considerations, it becomes problematic
near close-packed orientations ($\theta \! = \! 0^\circ$), where the kinks must be thermally
activated.  Therefore, we make use of exact results to assess in several ways how small $\theta$
can be before the simple expression becomes unreliable.  In Section III, we present three
general results for island stiffness that are valid in the experimentally-relevant low-$T$ limit
when configurational considerations dominate the thermodynamics. We show that the line tension
cannot be [re]generated from the stiffness and that the stiffness can have full 6-fold symmetry
even though the substrate and the line tension have just 3-fold symmetry. Accounting for such
3-fold symmetry with a lattice-gas model on a hexagonal grid requires an extension from the
conventional parametrization; we posit an {\em orientation-dependent} interactions between 3
atoms at the apexes of an equilateral triangle with NN legs.  In Section IV, we compare our
results to experiments on Ag and Cu\{111\} 
surfaces, showing that Eq.~(\ref{eq:stiffness}) provides a good approximation and, thus, that
NNN interactions are much less important than on Cu\{001\}.  The final section offers a
concluding discussion.  Three appendices give detailed calculations of the leading correction of
the low-$T$ expansion, of explicit analytic and numerical results based on Zia's exact implicit
solution, \cite{zia} and of Eq.~(\ref{eq:stiffness}) as the low-$T$ limit of Zia's solution.

\section{Ising Expansion on a Triangular Lattice}
\subsection{Recap of Results for Square Lattice}

\begin{figure}
\includegraphics[width=8.5 cm]{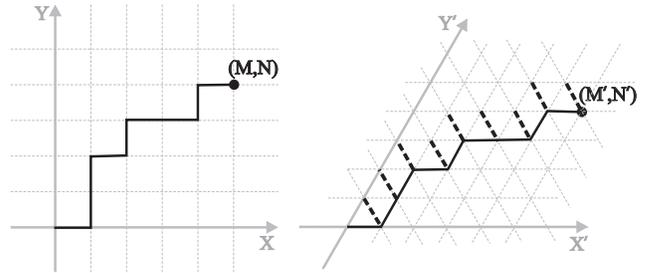}
\caption {There is a one-to-one correspondence between the shortest-distance steps connecting 
points on a square lattice [(a), left panel] and the shortest distance steps connecting points
on a 
triangular lattice [(b), right panel].  This figure shows two corresponding steps.
Analogous to the $M + N$ broken bonds oriented at $0^\circ$ and $90^\circ$ on a square lattice, 
there are $M^{\prime} + N^{\prime}$ broken bonds oriented at $0^\circ$ and at $60^\circ$ on a
triangular lattice.  
However, there are another $M^{\prime} + N^{\prime}$ broken bonds oriented at $120^\circ$ on a
triangular lattice.
}
{\label{fig:brokenBonds}}
\end{figure}

The orientation dependence of $\beta$ on \{111\} surfaces can be determined by
first calculating the free energy $F$ of a single step oriented at 
a fixed angle $\theta$.
To approximate this, we perform a low-temperature Ising expansion  
of the partition function, similar 
to the one used by Rottman and Wortis.\cite{rottman}  They
considered a step on a square 
lattice with one end fixed to the origin and the other end, a distance $L$ away, fixed to
the point ($M \equiv L \cos \theta$, $N \equiv L \sin \theta$).  Such a step is shown in 
Fig.~\ref{fig:brokenBonds}a.  The single-layer island (or compact adatom-filled region) is in
the lower region, separated by the step edge --- which is drawn as a bold solid line --- from
the upper part of the figure, representing the ``plain" region. 
They found the step energies associated
with the broken bonds of the step-edge to be
\begin{equation}
  \label{eq:energy_square}
  E_n = \varepsilon (M + N + 2 n), ~n= 0,1,2,\ldots ,
\end{equation}
where $\varepsilon$, sometimes\cite{Ising45} called the ``Ising parameter," is the bonding
energy associated with the ``severed half" of the NN lattice-gas bond:  Since  
the NN lattice-gas energy $\epsilon_1$ is attractive (negative), and half of it is attributed to
the atom on each end, it ``costs" a positive energy $\varepsilon = -\frac{1}{2}\epsilon_1$ for
each step-edge atom.  Because longer 
steps require more step-edge atoms, the step energy is only a function 
of the step length: $M + N + 2 n$. 
Thus, $E_0$ corresponds to 
the shortest possible step.  
To increase the length of this 
step, two more step-edge links -- corresponding to one more step-edge atom -- 
must be added, one going away from the 
fixed endpoint and one going toward it.  Because this corresponds
to two more broken bonds, in general, 
$E_{n+1} - E_n \equiv \Delta E = 2 \varepsilon$.  With these energies, we can write down the
partition function $Z_{\theta}$, assuming $\theta$ is fixed but $L$ is large enough so that
integer values of $M$ and $N$ can be found:
\begin{eqnarray}
  \label{eq:partitionFun}
  Z_{\theta} = g_{M,N}(0)e^{-E_0/k T}+g_{M,N}(1)e^{-E_1/k T}+... 
\end{eqnarray} 
where $g_{M,N}(n)$ corresponds to the number of ways a step of 
length $M + N + 2n$ can 
be arranged between the two endpoints.  

For low temperatures, 
only the first term in Eq.~(\ref{eq:partitionFun}) need be considered (Appendix I provides the
leading 
correction term, which gives a correction of order $\exp(-2\varepsilon/k_BT)$). 
To lowest order, then, $F$ is 
\begin{eqnarray}
  \label{eq:freeEnergy0}
  F \approx E_0 - k_BT \ln {M \! + \! N \choose M}
\end{eqnarray}
where we have inserted the value of $g_{M,N}(0)$ obtained from a simple combinatorial
analysis.\cite{Cahn,rottman} After taking the thermodynamic limit ($M$, $N$ $\gg$ $1$) and using
Stirling's approximation, $F$ becomes
\begin{eqnarray}
   \label{eq:freeEnergy}
    F \! \approx \!  E_0\!  -\!  k_BT \left[ (M\! +\! N) \ln (M\! +\! N)\! -\! M \ln M \! -\! N
\ln N \right]
\end{eqnarray}

\subsection{Triangular Lattice Step Energy} 

In order to extend this result to a step on a triangular
lattice, we need to make a few minor adjustments.
First, we introduce  
a linear operator $\cal L$ that  
transforms the coordinates of a point on a square lattice ($M$,$N$)
to those on a triangular lattice ($M^{\prime}$,$N^{\prime}$); cf.\ Fig.~\ref{fig:brokenBonds}b. 
This  
operator finds the position of a point in a coordinate 
system whose positive y-axis is bent at $60^\circ$ with respect to the 
positive x-axis:
\begin{eqnarray}
  \label{eq:LDef}
  \begin{pmatrix}M^{\prime} \\ N^{\prime} \end{pmatrix} &=& {\cal L}
  \begin{pmatrix}M  \\ N \end{pmatrix} \nonumber \\ 
  &=& \begin{pmatrix} 1 & -1/\sqrt{3} \\ 0 & 2/\sqrt{3}\end{pmatrix} 
      \begin{pmatrix}M \\ N \end{pmatrix}.
\end{eqnarray}

With the aid of $\cal L$, we can see how $E_0$ changes on a triangular lattice.  
To begin, we imagine a step in the first sextant (from $0^\circ$ to $60^\circ$ 
degrees in the plane) starting at (0,0) and ending at ($M'$, $N'$).  Such a step 
is shown in Fig. \ref{fig:brokenBonds}b.  As before, the bold solid line represents the step
edge with the bottom region a layer higher than the top (or, alternatively phrased, it 
separates the upper, adatom-free region from the lower, adatom-filled region). 
The 
broken bonds required to form the step will have only three orientations: 
$0^\circ$, $60^\circ$, and $120^\circ$.  
If we consider the shortest step between the two points (corresponding to energy $E_0$), then  
there will be exactly $M^{\prime} + N^{\prime}$ broken bonds oriented at $0^\circ$ and
$60^\circ$ (these 
bonds are analogous to those oriented at $0^\circ$ and $90^\circ$ on a square lattice).
There will be another $M' + N'$ broken bonds oriented at $120^\circ$ (drawn as 
bold, dashed lines in Fig. \ref{fig:brokenBonds}).  In total, there will be $2(M' + N')$ broken
bonds.  
Since $\varepsilon$ is the energy of these severed bonds, $E_0^\vartriangle = 2\varepsilon
(M^{\prime} + N^{\prime})$.  Thus, the energy is proportional to the 
step length, as was the case on a square lattice.    
Using $\cal L$ to 
write $M^{\prime}$ and $N^{\prime}$ in terms of $M$ and $N$ gives         
\begin{equation}
  \label{eq:E_0Final}
  E_0^\vartriangle = 2\varepsilon \left(M + \frac{ N}{\sqrt{3}} \right)
  = 2\varepsilon L  \left(\cos \theta + \frac{\sin \theta}{\sqrt{3}} \right) .
\end{equation}

\subsection{Triangular Lattice Step Degeneracy}

Next we consider 
the degeneracy factors $g^\vartriangle(n)$ on a triangular lattice.  
For the ground state $g^\vartriangle(0)$ there is a one-to-one correspondence between 
the shortest distance steps connecting two points on a square 
lattice and the corresponding steps on a triangular lattice
(see Fig. \ref{fig:brokenBonds}). 
Therefore, we know that the degeneracy factor $g^\vartriangle_{M^{\prime},N^{\prime}}(0)$ 
for steps of energy $E_0^\vartriangle$ on a triangular lattice 
must be identical to $g_{M^{\prime},N^{\prime}}(0)$ implicit in Eq.~(\ref{eq:freeEnergy0})!
More precisely, if we assume the point ($M$,$N$) is 
in the first quadrant, and ($M^{\prime}$, $N^{\prime}$) is in the first 
sextant, 
then on a square lattice, shortest-distance step-links are oriented at either 
$0^\circ$ or $90^\circ$, 
whereas on a triangular lattice they are oriented at either $0^\circ$ or $60^\circ$ 
(i.e. in the first sextant, 
the shortest path cannot have links oriented at 
$120^\circ$).  In both cases, the individual step-links can only be oriented 
in one of two directions and, therefore, besides the transformation between coordinates,
 the total number of path arrangements is the same.  

Using Eq.~(\ref{eq:freeEnergy0}) and Stirling's approximation, we find the low-temperature
free energy (Appendix I provides the lowest-order correction):
\begin{eqnarray}
  \label{eq:freeEnergySimp}
  \lefteqn{F \approx E_0^\vartriangle - k_BT \, \ln [g_{M',N'}(0)] \approx  E_0^\vartriangle -
k_BT \times} \nonumber \\
   &&   \times \left[ (M' \! + \! N') \ln (M'\! + \! N') 
     - M' \ln M' - N' \ln N' \right] . 
\end{eqnarray}

Alternatively, we can transform Eq.~(\ref{eq:freeEnergy}) for the square lattice to the
triangular lattice by just replacing $N/M \equiv \tan \theta$ with
$(2 \tan \theta)/(\sqrt{3} - \tan \theta)$.  (This ratio is just $N^{\prime}/M^{\prime}$, so it
might be termed $\tan \theta^{\prime}$.)  [We must also make a simple (and ultimately
inconsequential) change to $E_0$.]

We now take the thermodynamic limit ($M^{\prime}$, $N^{\prime}$ $\gg$ $1$) and write $M'$ and
$N'$ in terms of 
$M \equiv L \cos \theta$ and $N \equiv L \sin \theta$ via Eq.~(\ref{eq:LDef}). Then  dividing by
$L$ and defining\cite{A92note}
\begin{eqnarray}
  \label{eq:etas}
\eta_\pm(\theta) = \cos \theta \pm \frac{\sin \theta}{\sqrt{3}},  \qquad
\eta_0(\theta) =  \frac{2}{\sqrt{3}}\sin \theta ,
\end{eqnarray}
all non-negative in the first sextant, we straightforwardly find the step-edge line tension (or
free-energy per unit length\cite{IS}) $\beta(\theta)$: 
\begin{eqnarray}
  \label{eq:lineTensionTheta}
  a_\parallel\beta(\theta) = 2\varepsilon \eta_+(\theta) - k_B T \left[ s_+(\theta) - 
s_-(\theta)
    - s_0(\theta)\right],
\end{eqnarray}
where $a_\parallel$ is the nearest-neighbor spacing and 
\begin{eqnarray}
  \label{eq:Tpm0}
  s_i(\theta) = \eta_i(\theta) \ln \eta_i(\theta), \quad i=+,0,- \; .
\end{eqnarray}
For the special case of the maximally kinked orientation, Eqs.~(\ref{eq:etas}) --
(\ref{eq:Tpm0}) reduce to
\begin{eqnarray}
  \label{eq:SGI}
  a_k\beta(30^\circ) = 2\varepsilon - k_BT \ln 2,
\end{eqnarray}
\noindent where $a_k = (\sqrt{3}/2)a_\parallel$ for the \{111\} surface.  This result for the
maximally kinked case (including steps at $\theta\! =\! 45^\circ$ on a square lattice) was
derived earlier from a direct examination of entropy.\cite{SGI99}

For specificity, we recall some established results.  For a hexagonal lattice with just
nearest-neighbor attractions, the critical temperature $T_c$ is long known:\cite{Wan45}
\begin{eqnarray}
  \label{eq:hexTc}
  k_BT_c = 2\varepsilon/\ln 3 \approx 1.82 \varepsilon \ .
\end{eqnarray}
\noindent From the equilibrium shape of islands over a broad temperature range, Giesen {\it et
al.}\cite{Ising45}
deduced that the free energy per lattice spacing in the maximally kinked directions is 0.27
$\pm$ 0.03eV 
on Cu\{111\} and slightly smaller, 0.25 $\pm$ 0.03eV, on Ag\{111\}.  Combining these results
with Eq.~(\ref{eq:SGI}), 
we find $\varepsilon$ is 0.126eV on 
Cu\{111\} and 0.117eV on Ag\{111\}.  In both cases, then, room temperature is somewhere between
$T_c/9$ and $T_c/8$.

\subsection{Main Result: Simple Expression for Low-T Stiffness}

As shown just above, the step-stiffness $\tilde{\beta}=\beta(\theta) +  
\beta^{\prime\prime}(\theta)$
computed from Eq.~(\ref{eq:lineTensionTheta}) depends to leading order only on the combinatoric
entropy terms $s_0$ and $s_\pm$ of Eqs.~(\ref{eq:lineTensionTheta}) and(\ref{eq:Tpm0}).  Hence, 
\begin{eqnarray}
  \label{eq:t_i}
 \frac{d^2 s_i}{d \theta^2} \equiv s_i^{''} 
    &=& - \eta_i \ln \eta_i + \frac{{\eta_i^{'}}^2-\eta_i^2}{\eta_i},  
\end{eqnarray}
so that\cite{thetanote}
\begin{eqnarray}
  \label{eq:dt_i}
 \tilde{s}_i &\equiv& s_i + s_i^{''} 
  = \frac{{\eta_i^{'}}^2-\eta_i^2}{\eta_i} .
\end{eqnarray}
With this notation, the reduced stiffness is
\begin{eqnarray}
  \label{eq:stiffnesstilde}
 \frac{\tilde{\beta}a_\parallel}{k_B T} 
    &=& \tilde{s}_0 + \tilde{s}_- - \tilde{s}_+  ,
\end{eqnarray} 
where 
\begin{eqnarray}
  \label{eq:tilde{t}}
  \tilde{s}_0 &=& \frac{2}{\sqrt{3}} \frac{\cos 2 \theta}{\sin \theta} , \\
  \tilde{s}_\pm &=& \frac{-2 \cos 2 \theta \mp 2 \sqrt{3} \sin 2 \theta} 
   { 3 \cos \theta \pm \sqrt{3} \sin \theta  }.
\end{eqnarray}
Adding these terms together gives our \emph{main} result -- a remarkably simple form for the
reduced stiffness in the low-temperature ($T \ll \varepsilon/k_B$) limit:
\begin{eqnarray}
  \label{eq:stiffness}
  \frac{k_B T}{\tilde{\beta}a_\parallel} = \frac{\sin(3 \theta)}{2 \sqrt{3}} = \frac{3m - m^3}{2
\sqrt{3}(1+m^2)^{3/2}},
\end{eqnarray}
\noindent where $m \equiv \tan \theta$.

\subsection{Synopsis of Exact Results and Application to Range of Breakdown of Low-T Limit Near
$\theta$=0} 
\begin{figure}[t]
 \includegraphics[angle=0, width=8.5 cm]{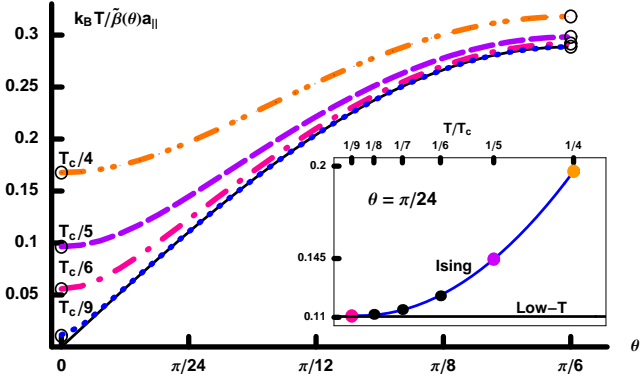}
 \caption{(Color online) As the temperature drops close to $T_c/9$ (just below room temperature
for Cu and Ag\{111\} surfaces), the numerical evaluation of the exact stiffness\cite{zia}
approaches the solid line representing the low-temperature approximation given in
Eq.~(\ref{eq:stiffness}). The small circles indicate evaluations using the exact results of
Eqs.~(\ref{eq:highsym2}) and (\ref{eq:highsym4}).  At $T_c/9$, when $\theta$ decreases, the
exact 
solution begins to deviate from the approximation when its curvature changes sign near $\theta
\approx  \pi/100 = 1.8^\circ$.  The scale here is linear, in contrast to the logarithmic scale
of Fig.~2 of Ref.~\onlinecite{dieluweit02}.  The inset shows more fully how the exact stiffness
approaches the low-temperature limit for the particular azimuthal angle $\theta \! = \! \pi/24
\! = \! 7.5^\circ$.
}
\label{fig:stiffnessCompare2}
\end{figure}

To test how low the temperature should be for Eq.~(\ref{eq:stiffness}) to be a good
approximation, we compare it to a numerical evaluation of the exact implicit solution of the
Ising 
model.  The derivation of this solution, outlined by Zia,\cite{zia} 
gives a $6^{th}$ order equation for $\beta(\theta)$.  In essence, after conversion to our
notation, his key result for the step free energy $\beta$ is given by\cite{zianote} 
\begin{eqnarray}
  \label{eq:betazia}
 \frac{\beta a_\parallel}{k_B T} = \eta_0(\theta)\psi_1(\theta,T/T_c) +
\eta_-(\theta)\psi_2(\theta,T/T_c), 
\end{eqnarray}
\noindent where the $\psi$'s are the solutions of the pair of simultaneous equations for the
angular constraint,
\begin{eqnarray}
  \label{eq:psithetzia}
 \frac{\sinh \psi_2 -\sinh(\psi_1 \! -\! \psi_2)}{\sinh \psi_1 +\sinh(\psi_1 \! -\! \psi_2)} =
\frac{\eta_-}{\eta_0} =  \frac{\sqrt{3}\cot \theta -\! 1}{2},
\end{eqnarray}
and the thermal constraint,
\begin{eqnarray}
\cosh \psi_1 \! +\! \cosh \psi_2 \! +\! \cosh (\psi_1 \! -\! \psi_2) \! =\!  f(z)\! \equiv \!
\frac{1+3z^2}{2(z-z^2)}, \quad
\label{eq:psiTzia}
\end{eqnarray}
\noindent where $z \equiv \exp (-2\varepsilon/k_BT) = 3^{-T_c/T}$, via Eq.~(\ref{eq:hexTc}). 
The ratio $\eta_-/\eta_0$ of Eq.~(\ref{eq:psithetzia}) is a monotonically decreasing function
which  
is $\infty$ at $\theta$=0$^\circ$, 1 at $\theta$=30$^\circ$, and 0 at $\theta$=60$^\circ$.  
 
In these high-symmetry directions, Eqs.~(\ref{eq:psithetzia}) and 
(\ref{eq:psiTzia}) yield analytic solutions for $\beta$ and 
$\tilde{\beta}$:
\begin{eqnarray}
 \frac{\beta(0) a_\parallel}{k_B T} &=& 2 \cosh^{-1}\left( \frac{-1+\sqrt{3 + 2 f}}{2}\right)
\label{eq:highsym1}\\
 \frac{\tilde{\beta}(0) a_\parallel}{k_B T} &=& \frac{2}{3} \frac{ \sqrt{2(3+2f)(f-\sqrt{3+2f})}
}{ \sqrt{3+2f}-1 }\label{eq:highsym2}\\
 \frac{\beta(\raisebox{0.5ex}{$\pi$} \!/\raisebox{-0.4ex}{\small 6}) a_\parallel}{k_B T} &=&
\frac{2}{\sqrt{3}}\cosh^{-1}\left(\frac{f-1}{2}\right) \label{eq:highsym3} \\ 
 \frac{\tilde{\beta}(\raisebox{0.5ex}{$\pi$} \!/\raisebox{-0.4ex}{\small 6}) a_\parallel}{k_B T}
&=& \frac{2\sqrt{3(f-3)(f+1)}}{f+3} \label{eq:highsym4}.
\end{eqnarray}
Details are provided in Appendix II. Akutsu and Akutsu \cite{AA99} also derived these equations,
in different notation \cite{AAnote} and from the more formal perspective of the imaginary
path-weight method.
Symmetry dictates that the solution
at $\theta=60^\circ=\pi/3$ is the same solution as that at $\theta=0^\circ$.   
Furthermore, at $T = T_c$, $f(z)=3$, so Eqs.~(\ref{eq:highsym1})-(\ref{eq:highsym4}) all go to
$0$, as expected.
\begin{figure}[b]
\includegraphics[width=8.5 cm] {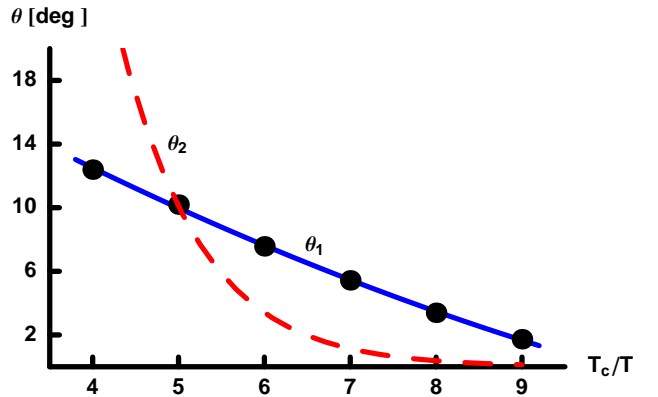}
\caption {(Color online) 
Two estimates for the critical angle $\theta_c$, below which the approximation given in
Eq.~(\ref{eq:stiffness}) begins to fail, as a function of $T_c/T$.  The black dots connected by
the solid, blue line represent the first estimate, 
defined to be the angle $\theta_1$ at which the curvature of the numerically determined inverse
stiffness changes 
sign.  The dashed, red line represents the second estimate, $\theta_2 =
\cot^{-1}[(4f-1)/(50\sqrt{3})]$.
At angles below $\theta_c$, the three theorems of Section \ref{sec:3thrm} break down, and higher
order terms are required in the expansion of the step partition function.  At temperatures
between $T_c/9$ and $T_c/8$ (roughly 
room temperature for Cu and Ag\{111\} surfaces), 
$\theta_c$ is on the order of a few degrees.   
}
\label{fig:thetac}
\end{figure} 

To find $\tilde{\beta}$ in general directions, we solve Eqs.~(\ref{eq:psithetzia}) and
(\ref{eq:psiTzia}) [or, equivalently, Eq.~(\ref{eq:stiffgensimp})]
numerically.  As Fig.~\ref{fig:stiffnessCompare2} shows, once $T$ decreases to nearly $T_c/9$, 
Eq.~(\ref{eq:stiffness}) more or less coincides with 
the exact numerical solution for the stiffness.  At such low 
temperatures (compared to $T_c$), the 
approximation only fails below some very small, temperature-sensitive critical angle $\theta_c$. 
Although 
it might seem easy to determine this angle by eye, estimating it  
quantitatively turns out to be a subtle and somewhat ambiguous task.  
We discuss two possible estimation techniques below.

In the first approach, we estimate $\theta_c$ to be the angle $\theta_1$ 
at which the {\it curvature} of the 
exact solution changes sign.  The points on the solid curve in Fig.~\ref{fig:thetac} 
show $\theta_1$ at several temperatures ranging 
from $T_c/9$ to $T_c/4$.  At temperatures near and 
above $T_c/4$, $\theta_1$ does not reliably estimate $\theta_c$ because there 
is a sizable curvature-independent difference between the exact 
solution and the approximation given in Eq.~(\ref{eq:stiffness}) evident even at $\theta =
30^\circ$ 
(see Fig.~\ref{fig:stiffnessCompare2}).  On the other hand, as the  
temperature dips below $T_c/5$,  
this difference fades, and the use of $\theta_1$ to estimate 
$\theta_c$ becomes ever more precise.  

A second, more fundamental way to estimate $\theta_c$ comes from an examination of 
the assumptions required to derive the simple expression for the low-T limit
Eq.~(\ref{eq:stiffness}) directly from the exact solutions Eqs.~(\ref{eq:psithetzia}) and 
(\ref{eq:psiTzia}).  In Appendix III we show that to do so $\theta$ must be greater
than some $\theta_2$ satisfying
\begin{eqnarray}
\cot\theta_2 \ll \left( \frac{4 f - 1}{\sqrt{3}}  \right).
\label{eq:thetac}
\end{eqnarray}
To give definite meaning to this inequality, we estimate $\theta_c$ directly from 
Fig.~\ref{fig:stiffnessCompare2}
at a 
single temperature, say $T_c/5$.  At that temperature, $\theta_c$  
is nearly $10^\circ$.  If $\theta_2$ is to accurately represent $\theta_c$, it should 
also be around $10^\circ$ at $T_c/5$.  We enforce this by interpreting    
the `$\ll$' in Eq.~(\ref{eq:thetac}) to mean `$=1/50$.'  The dashed (red online) curve in 
Fig.~\ref{fig:thetac} shows the resulting $\theta_2$ as a function of temperature.  
\begin{figure}[b]
\includegraphics[width=8.5 cm] {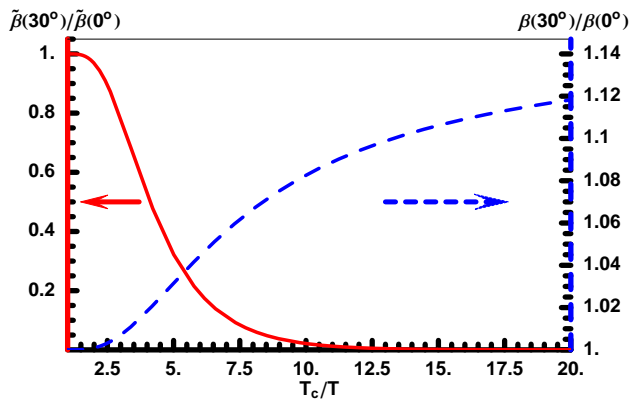}
\caption {(Color online)
Ratio of the stiffness [solid red curve, left ordinate] and the free energy per length [dashed
blue curve, right axis] for edges oriented in the maximally zig-zagged ($\theta = 30^\circ$) and
close-packed ($\theta = 0^\circ$) directions, based on taking the ratios of
Eqs.~(\ref{eq:highsym4}) and (\ref{eq:highsym2}) and of  
Eqs.~(\ref{eq:highsym3}) and (\ref{eq:highsym1}), respectively. The line-tension ratio increases
slowly but monotonically to the $T$=0 limit $2/\sqrt{3} \approx 1.15$.  In contrast, the
stiffness ratio plummets toward 0, the value predicted by Eq.(\ref{eq:stiffness}), providing an
indicator how low $T$ must be for this simple low-$T$ formula to be a good approximation at all
angles.
}
{\label{fig:ratio0-30}}
\end{figure} 
Clearly $\theta_1$ and $\theta_2$ are very different estimates for $\theta_c$.  While 
$\theta_2$ is reliable at all temperatures (unlike $\theta_1$), it is less 
precise than $\theta_1$ at lower temperatures.
A combination of $\theta_1$ and $\theta_2$ is therefore the best estimate for $\theta_c$, 
being closer to $\theta_1$ at lower temperatures, and closer to $\theta_2$ at 
higher temperatures.    

In essence, $\theta_c$ is no more than 
a few degrees between $T_c/9$ and $T_c/8$, regardless of which estimation technique is used. 
We therefore reach the practical  
conclusion that Eq.~(\ref{eq:stiffness}) is valid for almost all angles at temperatures near and
below 
$T_c/8$, which fortunately happens to be around room temperature for Cu and Ag\{111\}.  

Finally, we emphasize that $\tilde{\beta}$ varies significantly with angle, especially at lower
temperatures (where the equilibrium crystal shape (ECS)
 is hexagonal rather than circular).  If one wants to approximate $\tilde{\beta}$ as isotropic
rather than using Eq.~(\ref{eq:stiffness}), one should not pick 
its value in the close-packed direction (viz.\ $\theta = 0^\circ$); Fig.~\ref{fig:ratio0-30}
provides stunning evidence of this conclusion.  
From Eq.~(\ref{eq:stiffness}) we also see that at low-temperatures 
the stiffness actually \emph{increases} linearly with temperature.  This contrasts with its
behavior at high temperatures, where $\tilde{\beta}$ must ultimately decrease as the ECS becomes
more nearly circular and the steps fluctuate more easily.

\section{General Results for Stiffness in Lattice-gas Models in Low-Temperature Approximation}
\label{sec:3thrm}
In this section we present three theorems that are valid under two conditions: First, the energy
term in the free energy must be a linear combination of $\cos \theta$ and $\sin \theta$.  From
Eq.~(\ref{eq:E_0Final}) and [implicitly] Eq.~(\ref{eq:freeEnergy0}) we see that this property
holds true in general for lattice-gas models, even when considering next-nearest neighbors and
beyond.\cite{Herring51}  Second, the temperature must be low enough so that the entropy is
adequately approximated by the contribution of the lowest order term, $k_B \ln g(0)$. This
entropic contribution is due exclusively to geometry or combinatorics of arranging the fixed
number of kinks forced by azimuthal misorientation.  Hence, it must vanish near close-packed
directions (0$^\circ$ and 60$^\circ$ in the first sextant).  For angles sufficiently close to
these directions, in our case less than $\theta_c$,
 the leading term becomes dominated by higher-order terms, and the three results no longer
apply.  

\subsection{No Contribution from Energy to Lowest-Order Stiffness (LOS)}
 The first theorem is a remarkable consequence of the first condition, that the energy term in
the free energy is a linear combination of $\cos \theta$ and $\sin \theta$.  Since the stiffness
$\tilde{\beta}(\theta)\equiv \beta(\theta) + \beta^{\prime\prime}(\theta)$ and since
$\cos^{\prime\prime} \theta = -\cos \theta$ and $\sin^{\prime\prime} \theta = -\sin \theta$, we
see that {\it the lattice-gas energy makes no contribution whatsoever to the low-T limit of
reduced stiffness}, as shown explicitly for square lattices long ago.\cite{rottman,Cahn}  Thus,
we retrieve the result that the leading term in a low-temperature expansion of the reduced
stiffness $\tilde{\beta}(\theta)/k_BT$ depends only on $g(0)$, which is determined solely by
geometric (combinatoric) properties.  Of course, higher-order terms (which are crucial near
close-packed directions) do have weightings of the various configurations that depend on
Boltzmann factors involving the characteristic lattice-gas energies.  Furthermore,
next-nearest-neighbor interactions can (at least partially) lift the $g(0)$-fold degeneracy of
the lowest energy paths.\cite{tjs001}

\subsection{Step Line Tension Not Extractable from LOS}
An important corollary is that from the stiffness it is impossible to retrieve the energetic
part of the step free energy, the major component of $\beta(\theta)$ at lower temperatures when
the islands are non-circular!  Thus, contrary to a proposed method of data analysis,\cite{OSDF}
one cannot regenerate $\beta(\theta)$ from $\tilde{\beta}(\theta)$ by fitting the stiffness to a
simple functional form and then integrating twice.  In this framework, the linear coefficients
of $\cos \theta$ and $\sin \theta$ can be viewed as the two integration constants associated
with integrating a second-order differential equation.\cite{Krug03}

\subsection{LOS on fcc\{111\} Has 6-fold Symmetry} 
\subsubsection{1. General Argument}
Another important result is that the leading term in the stiffness at low temperature has the
full symmetry of the 2D net of binding sites rather than the possibly lower symmetry associated
with the full lattice.  Specifically, for the present problem of the \{111\} face of an fcc
crystal, the stiffness $\tilde{\beta}(\theta)$ to lowest order has the full 6-fold symmetry of
the top layer rather than the 3-fold symmetry due to symmetry breaking by the second layer.  In
contrast, the step energy of B-steps (\{111\} microfacets) differs from that of A-steps (\{100\}
microfacets), leading to islands with the shape of equiangular hexagons with rounded corners,
but with sides of alternating lengths (i.e.,\ ABABAB).  

To see the origin of the 6-fold symmetry of the stiffness, suppose without loss of generality
that steps in the $X'$ direction have energy $E_A$ per lattice spacing, so that those in the
$Y'$ direction have energy $E_B$. Furthermore, we must make the crucial assumption that any
corner energy is negligible.  Then all shortest paths to ($M',N'$) have the same energy $M'E_A +
N'E_B$, with degeneracy still $g_{M^{\prime},N^{\prime}}(0)$.  Thus, the free energy is $M'E_A +
N'E_B -k_BT \ln g_{M^{\prime},N^{\prime}}(0)$, while that of its mirror point (through the line
at $\theta = 30^\circ$) is  $N'E_A + M'E_B -k_BT \ln g_{N^{\prime},M^{\prime}}(0)$.  The crux of
the proof is that $g_{N^{\prime},M^{\prime}}(0)= g_{M^{\prime},N^{\prime}}(0)$.  Thus, while the
free energies at the pair of mirror points differ, the energy parts are obliterated when the
stiffness is computed (since $M'$ and $N'$ are linear combinations of $\cos \theta$ and $\sin
\theta$), leaving just the contribution from the entropies, which are the same to lowest order.  

\subsubsection{2. Orientation-Dependent 3-Atom Interaction}
Within lattice-gas models with only pair interactions, there is no obvious way to distinguish A
and B steps; the minimalist way to obtain different step energies for A and B steps within the
lattice-gas model is to invoke a non-pairwise 3-site ``trio" interaction associated with three
[occupied] sites forming an equilateral triangle with NN sides.
In contrast to the ones considered heretofore,\cite{triorev,TLE-Unertl} these novel trio
interactions must be {\it orientation-dependent}:  If the triangle points in one direction, say
up, the interaction energy is positive, while if it points in the opposite direction, it has the
opposite sign.  (Of course, there could be a standard orientation-independent 3-site term in the
Hamiltonian.  As in the analogous situation for squares, we expect that such a term would simply
shift the pair interactions, at least in the SOS approximation as discussed in
Ref.~\onlinecite{tjs001}.)  The contributions from such a symmetry-breaking interaction would
cancel in the interior of an island (in the 2D bulk), but would distinguish A and B edges. 
Specifically, each side of the equilateral triangle is associated with a link, so that 1/3 of
its strength can be attributed to each.  Each link has a triangle on both sides, one of each
orientation.  Hence, the difference between the energy per $a_\parallel$ of A and B steps is 1/3
the difference between the trio interactions in the two opposite orientations.

For the ground-state, minimum-number-of-links configurations, such a term will not lift the
degeneracy since each configuration has the same 1) number of horizontal ($X'$) links, 2) number
of right-tilted diagonal links ($Y'$), and 3) {\it difference} between the number of convex and
concave ``kinks" (i.e., bends).  Since this statement is not true for higher-energy
configurations, the 6-fold symmetry is not preserved at higher orders.  Nonetheless, at low $T$
it should be a decent approximation for the stiffness (much better than for the island shape).

Thus, our result that the breaking of 6-fold symmetry on an fcc \{111\} is much smaller for the
stiffness than for the free energy, is more general than the nearest-neighbor lattice gas model
which underlies Eqs.~(\ref{eq:lineTensionTheta}) -- (\ref{eq:Tpm0}) and the resulting
Eq.~(\ref{eq:stiffness}) derived below.  We reemphasize that the necessary assumptions are 1)
that the orientational dependence of the step energy be just a linear combination of $\sin
\theta$ and $\cos \theta$ and 2) that no interaction break the degeneracy of the shortest path
corresponding to orientation $\theta$.  As above, for angles near close-packed directions, the
higher-order terms become important at lower temperatures than for general directions.  This
feature is illustrated in Fig.~\ref{fig:stiffnessCompare2} and its associated formalism is given
in Appendices I and III.

\section{Comparison to Experiment}

 \begin{figure}[t]
 \includegraphics[angle=0, width=9 cm]{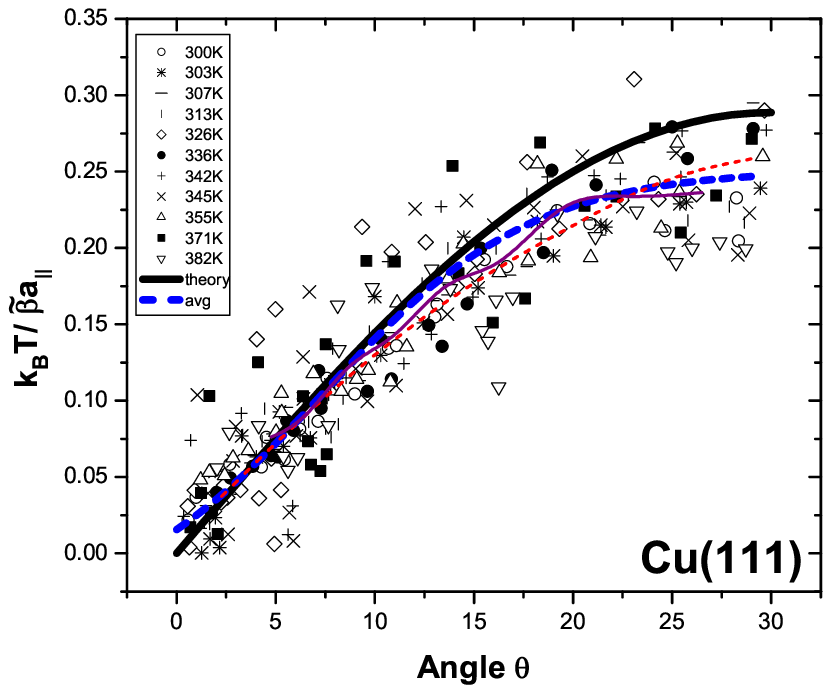}
 \includegraphics[angle=0, width=9 cm]{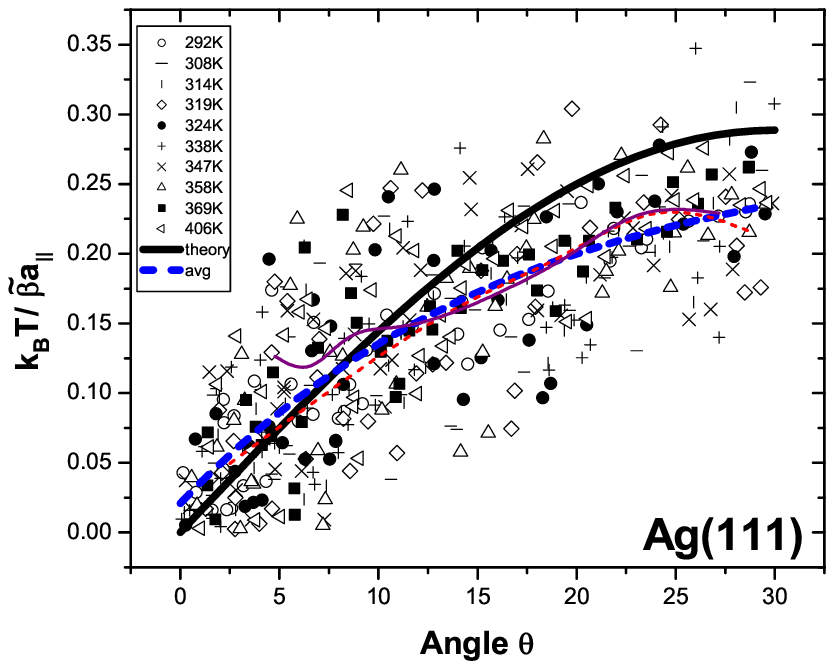}
 \caption{(Color online) 
A comparison between 
Eq.~(\ref{eq:stiffness}) and experiments on Cu and Ag\{111\}.  Eq.~(\ref{eq:stiffness}) appears
as a solid black line, while 
the average of the experimental data is a thick dashed blue line.  The agreement is reasonable
at all angles.
In either case the thin dashed red line is a [smoothed] average of the data for the given angle
while the thin solid purple line corresponds to the angle mirror-reflected through a radial at
30$^\circ$, i.e. at 60$^\circ \! -\! \theta$.
}
{\label{fig:giesen111}}
 \end{figure}

In Fig.~\ref{fig:giesen111} we compare Eq.~(\ref{eq:stiffness}) to measurements on Cu\{111\} and
Ag\{111\}.  The experimental data were derived from the equilibrium shape of 2D islands using
the method described in Ref.~\onlinecite{dieluweit02}.  The solid black line corresponds to 
Eq.~(\ref{eq:stiffness}), while the thick dashed blue line corresponds to the average of the
experimental measurements.  
Eq.~(\ref{eq:stiffness}) captures the overall trend and is satisfactory at most angles and
temperatures.  As 
expected, Eq.~(\ref{eq:stiffness}) somewhat overestimates $\tilde{\beta}$ near $\theta =
0^\circ$ (since the 
$T\! = \! 0$ singularity remains).  Furthermore, near $\theta = 30^\circ$
Eq.~(\ref{eq:stiffness}) somewhat underestimates the experimental $\tilde{\beta}$, but only by a
factor of 1/6 for Cu\{111\} and 1/4 for Ag\{111\}.  This is in striking contrast to the
analogous NN theory for Cu\{001\} near 45$^\circ$, which underestimates $\tilde{\beta}$ by a
factor of 4.
Finally, notice there is no clear temperature dependence in the measured data.  This is further
evidence that 
$\tilde{\beta}/k_BT$ is a constant at low-temperatures, as 
Eq.~(\ref{eq:stiffness}) suggests.

The agreement between theory and experiment 
is a pleasant surprise considering analogous 
comparisons made 
for Cu\{001\}\cite{dieluweit02} found $\tilde{\beta}$ to be four-times larger than the
theoretical value at large angles (near 
$\theta = 45^\circ$).
It was later shown\cite{ZP1,ZP2,tjs001} that this discrepancy could be partially accounted for
by considering 
next-nearest-neighbor (NNN) interactions (or right-triangle trio interactions, which turn out to
affect $\tilde{\beta}$ 
at low temperatures in the same way).  
Clearly, the success of Eq.~(\ref{eq:stiffness}) suggests that these 
interactions are less relevant for \{111\} surfaces.  This is reasonable because the ratio of
NNN distance  
to NN distance is smaller by a factor of $\sqrt{2/3}$ on a triangular lattice compared to on a
square lattice.  
Furthermore, in the close-packed direction ($\theta = 0^\circ$), for every broken NN-bond there
are only 
one and a half broken NNN bonds on a triangular lattice, compared to two broken 
NNN bonds on a square lattice.  These simple arguments help explain why   
NNN-interactions may 
increase $\tilde{\beta}$ by only $20$ to $30$\% on Cu/Ag\{111\}, as opposed to $400$\% on
Cu\{001\} surfaces.

\section{Concluding Discussion}

By generalizing the low-temperature expansion of the nearest-neighbor square lattice-gas (Ising)
model to a triangular lattice, we have found a remarkably simple 
formula for the orientation dependence of the \{111\} surface step stiffness.  This formula,
unlike its square 
lattice analog, fits experimental data well at general angles, suggesting that NNN-interactions
are relatively unimportant on \{111\} surfaces.  

To corroborate this picture and explain the success of Eq.~(\ref{eq:stiffness}), we are
currently using 
the VASP package\cite{VASP}
to perform first-principle calculations.  In particular,
we are examining the 
ratio of the NNN to NN interaction strength.  Preliminary results\cite{STK} suggest that this 
ratio is roughly an order of magnitude smaller on Cu\{111\} than on Cu\{001\}, and essentially
indistinguishable from zero.  This tentative finding is consistent with expectations from the
semiempirical embedded atom method, which predicts that indirect interactions are
insignificant/negligible between atoms sharing no common substrate atoms.\cite{TLE-Unertl}  We
are also calculating the difference in trio interactions between oppositely oriented triangle
configurations.

We expect that our formula, as well as the general 6-fold symmetry of the stiffness (except in
close-packed directions), should be broadly applicable to systems in which multisite or corner
energies are small and for which the bond energies are considerably higher than the measurement
temperature. Studies which ignore the 3-fold symmetry breaking on metallic fcc \{111\}
substrates, such as a recent investigation of nanoisland fluctuations on Pt\{111\},\cite{Sz04}
should be good representations.  Many recent investigations\cite{Danker,KK04} focus on the
larger asymmetry of the kinetic coefficient,\cite{GD04} taking the stiffness to be isotropic. 
In such cases, this stiffness should not be characterized by its value in close-packed
directions. 

\section*{APPENDIX I: LEADING TERM IN LOW-TEMPERATURE EXPANSION}   
\subsection{Review of Results for Square Lattice}
    In this appendix we discuss the lowest-order correction to the ground state entropy of the
step running from the origin to an arbitrary particular point.  First we review results for a
square lattice.  We can rewrite Eq.~(\ref{eq:partitionFun}) as
      \begin{eqnarray}
  \label{eq:partitionFun1}
  Z_{\theta} = g_{M,N}(0)e^{-E_0/k T}\left[1 + \frac{g_{M,N}(1)}{g_{M,N}(0)}e^{-\Delta E/k T} +
...\right].  
\end{eqnarray} 
\noindent Then, assuming the exponential is small, we have 
\begin{eqnarray}
  \label{eq:freeEnergy0a}
  F \approx E_0 - k_BT \left\{ \ln [g_{M,N}(0)]+\frac{g_{M,N}(1)}{g_{M,N}(0)} 
       e^{-2\varepsilon/k_BT}\right\}.  
\end{eqnarray}

\noindent A combinatorial analysis\cite{Cahn,rottman,tjsnote1} shows that 
\begin{eqnarray}
 \label{eq:g1}
   g_{M,N}(1) = {M\! +\! N \choose M\! -\! 1}(M\! +\! 1) 
   + {M\! +\! N \choose N\! -\! 1}(N\! +\! 1)
  \end{eqnarray}

\noindent Then Eq.~(\ref{eq:freeEnergy}) generalizes to 

\begin{eqnarray}
   \label{eq:freeEnergya}
    F &\approx& E_0 - k_BT \left[(M\! +\! N) \ln (M\! +\! N) - M \ln M - N \ln N
     \right. \nonumber \\
    && \left.   + e^{-2\varepsilon/k_BT} 
    \frac{M^3+N^3}{MN} \right]. 
\end{eqnarray}

\subsection{Results for Triangular Lattice}      
For the triangular lattice we find important differences from the square lattice for the
higher-order terms.  Specifically, we consider how $g(1)$ changes.  In contrast to $g(0)$, we
cannot 
simply replace $M$ and $N$ with $M^{\prime}$ and $N^{\prime}$.  There is no 
one-to-one correspondence between paths of energy 
$E_1$ on a square lattice and those of energy 
$E_1^\vartriangle$ 
on a triangular lattice.  This failed correspondence for higher terms follows from the
observation that $E_1^\vartriangle$-configurations are only one link 
longer than $E_0^\vartriangle$-steps, whereas $E_1$-configurations are two links longer 
than $E_0$-steps:  
$E_{n+1}^\vartriangle - E_n^\vartriangle \equiv \Delta E^\vartriangle = \varepsilon$, or
\begin{eqnarray}
  \label{eq:etri}
  E_n^\vartriangle = \varepsilon \left(\frac{2 N}{\sqrt{3}} + 2 M + n\right), \, n=0,1,2,...,
\end{eqnarray} 
\noindent Hence, we require a separate combinatorial analysis.  

 \begin{figure}[t]
 \includegraphics[width=6 cm]{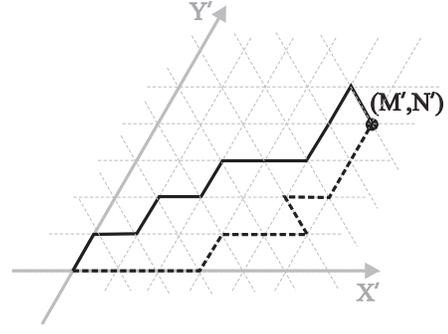}
 \caption {Two equivalent steps having energy $E_1^\vartriangle$.  The dashed 
step  
contains ($M^{\prime} + 1$) X-links, ($N^{\prime} - 1$) Y-links, and one $\Upsilon$-link, while
the 
solid step contains ($M^{\prime} - 1$) X-links, ($N^{\prime} + 1$) Y-links, and one
$\Upsilon$-link.}
{\label{fig:E1paths}}
\end{figure}
We imagine a step of energy $E_1$ in the first sextant.  Such a step 
(see Fig. \ref{fig:E1paths})
will have either: 
(1) ($M^{\prime}+1$) links oriented at $0^\circ$ (denoted ``X-links"), 
($N'-1$) links oriented at $60^\circ$ (denoted ``Y-links"), 
and one link oriented at $120^\circ$ (denoted ``$\Upsilon$-links"), or 
(2) ($M^{\prime}-1$) links oriented at $0^\circ$, 
($N'+1$) links oriented at $60^\circ$, 
and one link oriented at $-60^\circ$.  In the first case, 
the problem can be reworded as follows: how many ways to arrange an 
($M^{\prime} + N^{\prime} +1$)-lettered word with ($M^{\prime} + 1$) X's, ($N^{\prime}- 1$) Y's, 
and one $\Upsilon$.  In the second 
case, the problem is the same, only with $M$ and $N$ switched.
Thus, the solution of this 
traditional combinatorial problem gives the total number of 
next-to-shortest paths $g^\vartriangle(1)$:
\begin{eqnarray}
  \label{eq:g1tri}
  g^\vartriangle(1)
   = {M^{\prime} \! +\!  N^{\prime} \! +\!  1 \choose M^{\prime}\!  +\! 1}N^{\prime} + 
   {M^{\prime}\! +\!  N^{\prime} \! +\! 1 \choose N^{\prime} \! +\! 1}M^{\prime}.
\end{eqnarray}
\indent With $g^\vartriangle(0/1)$ and $E_n^\vartriangle$ in hand, we can write the 
low-temperature partition function 
expansion for a triangular lattice.  Using Eq.~(\ref{eq:freeEnergy0}) and expanding the
logarithm as in Eq.~(\ref{eq:freeEnergy0a}), we have
\begin{eqnarray}
  \label{eq:freeEnergyTri0}
  F &\approx& E_0^\vartriangle - k_BT \,  \left\{ \ln [g(0)]+
\frac{g^\vartriangle(1)}{g(0)} 
       e^{-\Delta E^\vartriangle/k_BT}\right\}.  
\end{eqnarray}
Taking the thermodynamic limit ($M^{\prime}$, $N^{\prime}$ $\gg$ $1$) and using Stirling's
approximation gives
\begin{eqnarray}
  \label{eq:freeEnergyTri}
\hspace{-2mm} F \! &\approx& \! E_0^\vartriangle \! - \! k_B T \left[ (M' \! + \! N')
\ln (M' \! + \! N') \! - \! M' \ln M' \! - \! N' \ln N'    
   \right. \nonumber \\
       \! &+& \! \left. e^{-\varepsilon/k_BT} \frac{M'^3 \! + \! N'^3 \! + \! M' N'^2 \! + \! N'
M'^2}{M'N'}   \right] \, . 
\end{eqnarray}
\noindent The pair of cross-factors in the last coefficient are absent in
Eq.~(\ref{eq:freeEnergya}) for the square lattice. 

The correction term becomes non-negligible when the final term in Eq.~(\ref{eq:freeEnergyTri})
becomes of order unity.  At low $T$ this occurs only near close-packed directions, so for small
values of $\theta$. In this regime, to lowest order in $\theta$, $N' = (2L/\sqrt{3})\sin \theta
\rightarrow 2L \theta /\sqrt{3}$ and $M' = L \cos \theta - N'/2 \rightarrow L$.  Then the
critical value of $\theta$ is 
\begin{eqnarray}
  \label{eq:crossover}
  \theta_c^{(\beta)} \approx \textstyle{\frac{\sqrt{3}}{2}}e^{-\varepsilon/k_BT} =
\textstyle{\frac{\sqrt{3}}{2}} z^{1/2}.  
\end{eqnarray}
\noindent 
Specifically, based on Eq.~(\ref{eq:crossover}) and using $\varepsilon \approx 0.12$ eV for
Cu\{111\}, we find that $\theta_c^{(\beta)}$ is 0.353$^\circ$, 3.18$^\circ$, and 5.51$^\circ$
for T/T$_c$ of 1/9, 1/5, and 1/4, respectively.  As clear from Fig.~\ref{fig:thetac}, this
criterion turns out to underestimate the values for $\theta_c$ obtained in Section II.D, 
mainly because Eq.~(\ref{eq:crossover}) was derived from 
an expression for $\beta(\theta)$ instead of $\tilde{\beta}(\theta)$ (which should depend more
sensitively on $\theta$).  
over the plotted thermal range.
\section*{APPENDIX II: EXACT FORMULAS FOR LINE TENSION AND STIFFNESS IN MIRROR DIRECTIONS}   
\subsection*{A. General results for all orientations}In this appendix, we derive
Eqs.~(\ref{eq:highsym1}) -- (\ref{eq:highsym4}) for the mirror-line directions $\theta=0^\circ$
and $\theta=30^\circ$ from Zia's implicit exact solution.\cite{zia}  

To begin, because $\tilde{\beta} = \beta + \beta''$ (where the prime 
represents differentiation with respect to $\theta$), it follows 
from Eq.~(\ref{eq:betazia}) that
\begin{eqnarray}
  \label{eq:stiffgen}
  \frac{\tilde{\beta} a_{||}}{k_B T} = 2 \eta_0' \psi_1' + 2 \eta_-' \psi_2' + \eta_0 \psi_1'' +
\eta_- \psi_2''.
\end{eqnarray}
We can simplify Eq.~(\ref{eq:stiffgen}) by finding 
relationships between the various 
derivatives of the $\psi$'s.  Differentiating 
Eq.~(\ref{eq:psiTzia}) with respect to $\theta$, regrouping, and using
Eq.~(\ref{eq:psithetzia}), we get 
\begin{eqnarray}
  \psi_1' \eta_0 + \psi_2' \eta_- &=& 0. 
\label{eq:dpsiTzia}
\end{eqnarray}
Differentiating again yields
\begin{eqnarray}
  \psi_1'' \eta_0 + \psi_2'' \eta_- + \psi_1' \eta_0' + \psi_2' \eta_-' &=& 0.
\label{eq:ddpsiTzia}
 \end{eqnarray}
Using Eq.~(\ref{eq:ddpsiTzia}), we rewrite the last part of Eq.~(\ref{eq:stiffgen}) (containing 
$\psi_1''$ and $\psi_2''$) in terms of just $\psi_1'$ and $\psi_2'$.  Then, using 
Eq.~(\ref{eq:dpsiTzia}) we eliminate $\psi_2'$ in favor of $\psi_1'$.  We are then left 
with an equation relating $\tilde{\beta}$ to only $\psi_1'$:
\begin{eqnarray}
  \label{eq:stiffgensimp}
  \frac{\tilde{\beta} a_{||}}{k_B T} = \left( \eta_0' - \eta_-' \frac{\eta_0}{\eta_-}  \right)
\psi_1'
  = \frac{2 \psi_1'}{\sqrt{3} \cos \theta - \sin \theta}.
\end{eqnarray}
For general angle, we must evaluate $\psi_1'$ numerically. However, for the two high-symmetry
directions we can obtain analytic results that allow us (with the aid of Eq.~(\ref{eq:betazia})
for $\beta$) to write explicit expressions for $\tilde{\beta}$, as presented in the next two
subsections.  

 \begin{figure}[b]
 \includegraphics[angle=0, width=8 cm]{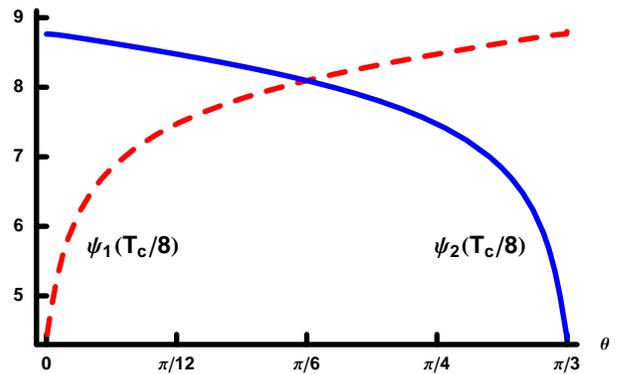}
 \caption{(Color online)
Numerical evaluation of $\psi_1$ (dashed red curve) and $\psi_2$ (solid blue curve) as functions
of angles at temperature $T_c/8$ equivalent to room temperature for the experimental systems Cu
and Ag \{111\}.  Note that the linear behavior near one limit and the divergent slope near value
zero 
at the other.  At higher temperatures the curves are qualitatively similar but progressively
smaller in magnitude.
}
{\label{fig:psiplot}}
 \end{figure}

\subsection*{B. Results for $\theta=0^\circ$}
At $\theta=0^\circ$, Eq.~(\ref{eq:betazia}) reduces to
\begin{eqnarray}
  \label{eq:betazia0}
  \frac{\beta a_{||}}{k_B T} = \psi_2(0),
\end{eqnarray}
\noindent assuming that $\psi_1(0)$ is finite.  
Furthermore, near $\theta=0^\circ$, Eq.~(\ref{eq:psithetzia}) can be inverted and the $\sinh$'s
combined to get
\begin{eqnarray}
\frac{\sinh(\psi_1-  \frac{1}{2} \psi_2) \cosh(\frac{1}{2}\psi_2)}{\sinh(\psi_2-\frac{1}{2}
\psi_1) \cosh(\frac{1}{2}\psi_1)}= \frac{2}{\sqrt{3}}~\theta. 
\label{eq:betazia2}
\end{eqnarray}
For Eq.~(\ref{eq:betazia2}) to hold at $\theta=0^\circ$,  
\begin{eqnarray}
  \label{eq:psi0}
\psi_2(0) = 2 \psi_1(0).  
\end{eqnarray}
Eq.~(\ref{eq:psiTzia}) therefore becomes:  
\begin{eqnarray}
2 \cosh \psi_1(0) + \cosh(2 \psi_1(0)) = f.
\label{eq:psi10}
\end{eqnarray}
Solving this for $\cosh \psi_1(0)$ 
and taking the positive root, we find:
\begin{eqnarray}
  \label{eq:psi10f}
  \cosh \psi_1(0) = \cosh (\textstyle{\frac{1}{2}}\psi_2(0))  = \textstyle{\frac{1}{2}} (-1 +
\sqrt{3+2f}),
\end{eqnarray}
consistent with the assumption of finite $\psi_1(0)$. Solving for $\psi_2(0)$ and 
combining with Eq.~(\ref{eq:betazia0}) yields 
Eq.~(\ref{eq:highsym1}).

Correspondingly for $\tilde{\beta}$, at $\theta=0^\circ$ Eq.~(\ref{eq:stiffgensimp}) becomes
\begin{eqnarray}
  \label{eq:stiffzia0simp}
   \frac{\tilde{\beta}(0) a_\parallel}{k_B T} &=& \frac{2}{\sqrt{3}} \psi_1'(0),
\end{eqnarray}
while Eq.~(\ref{eq:dpsiTzia}) becomes
\begin{eqnarray}
  \label{eq:psi2prime0}
  \psi_2'(0) = 0,
\end{eqnarray}
provided $\psi_1'(0)$ is finite.  
We obtain $\psi_1'(0)$ by differentiating Eq.~(\ref{eq:betazia2}) with respect to $\theta$ and
then setting 
$\theta=0^\circ$ so that Eqs.~(\ref{eq:psi0}) and (\ref{eq:psi2prime0}) apply.  This give  
\begin{eqnarray}
  \label{eq:dpsi1}
  \psi_1'(0) = \textstyle{\frac{1}{\sqrt{3}}} \tanh \psi_1(0) [1+2 \cosh \psi_1(0)].
\end{eqnarray}
By combining this with Eq.~(\ref{eq:psi10f}) for $\cosh \psi_1(0)$, we see that $\psi_1'(0)$ is 
indeed finite, as we earlier assumed.  Thus,     
Eq.~(\ref{eq:stiffzia0simp}) becomes Eq.~(\ref{eq:highsym2}), as desired.

\subsection*{C. Results for $\theta= 30^\circ$}

At $\theta=30^\circ=\pi/6$, Eq.~(\ref{eq:betazia}) becomes
\begin{eqnarray}
  \frac{\beta(\raisebox{0.4ex}{$\pi$} \!/ \raisebox{-0.3ex}{\footnotesize 6}) a_\parallel}{k_B
T} &=& \textstyle{\frac{1}{\sqrt{3}}} \left[\psi_1(\raisebox{0.5ex}{$\pi$}
\!/\raisebox{-0.4ex}{\small 6})+\psi_2'(\raisebox{0.5ex}{$\pi$} \!/\raisebox{-0.4ex}{\small
6})\right]. \label{eq:betazia30}
\end{eqnarray}
Furthermore, near $\theta=\pi/6$, $\eta_0/\eta_- \approx 1 + 2 \sqrt{3}~\Delta \theta$, 
where $\Delta \theta \equiv \theta - \pi/6$.  Inverting Eq.~(\ref{eq:psithetzia}) 
we therefore have
\begin{eqnarray}
  \label{eq:psithetzia30}
 \frac{\sinh \psi_1 +\sinh(\psi_1 - \psi_2)}{\sinh \psi_2 -\sinh(\psi_1 - \psi_2)} 
\approx 1 + 2 \sqrt{3} ~ \Delta \theta,
\end{eqnarray}
By inspection, at $\theta=\pi/6$ ($\Delta \theta = 0$), one solution to this equation is  
just 
\begin{eqnarray}
  \label{eq:psi30}
\psi_2(\raisebox{0.5ex}{$\pi$} \!/\raisebox{-0.4ex}{\small 6}) = \psi_1(\raisebox{0.5ex}{$\pi$}
\!/\raisebox{-0.4ex}{\small 6}).  
\end{eqnarray}
Plugging this result   
into Eq.~(\ref{eq:psiTzia}) and solving for $\psi_2(\pi/6)$ gives, 
\begin{eqnarray}
  \label{eq:psi230}
  \cosh \psi_2(\raisebox{0.5ex}{$\pi$} \!/\raisebox{-0.4ex}{\small 6}) = \frac{f-1}{2}.
\end{eqnarray}
Combining this with Eq.~(\ref{eq:betazia30}) (where we now know $\psi_1(\pi/6)=\psi_2(\pi/6)$)
 results in Eq.~(\ref{eq:highsym3}).

As for $\tilde{\beta}$, at $\theta=\pi/6$,  Eq.~(\ref{eq:stiffgensimp}) becomes
\begin{eqnarray}
  \label{eq:stiffgensimp30}
  \frac{\tilde{\beta} a_{||}}{k_B T} = 2 \psi_1'(\raisebox{0.5ex}{$\pi$}
\!/\raisebox{-0.4ex}{\small 6}),
\end{eqnarray}
while Eq.~(\ref{eq:dpsiTzia}) becomes
\begin{eqnarray}
\psi_1'(\raisebox{0.5ex}{$\pi$} \!/\raisebox{-0.4ex}{\small 6})=
-\psi_2'(\raisebox{0.5ex}{$\pi$} \!/\raisebox{-0.4ex}{\small 6}). \label{eq:dpsiTzia30}  
\end{eqnarray}
Like before, we can find $\psi_1'(\pi/6)$ by differentiating Eq.~(\ref{eq:psithetzia30}) with 
respect to $\theta$.  Taking the result and setting
 $\theta=\pi/6$, so that Eqs.~(\ref{eq:psi30}) and (\ref{eq:dpsiTzia30}) apply,  
gives
\begin{eqnarray}
  \label{eq:dpsi130}
  \psi_1'(\raisebox{0.5ex}{$\pi$} \!/\raisebox{-0.4ex}{\small 6})=\frac{\sqrt{3} \sinh
\psi_1}{\cosh \psi_1+2}.
\end{eqnarray}
Finally, we combine this result with 
Eq.~(\ref{eq:psi230}) for $\cosh\psi_2(\raisebox{0.5ex}{$\pi$} \!/\raisebox{-0.4ex}{\small 6}) =
\cosh\psi_1(\raisebox{0.5ex}{$\pi$} \!/\raisebox{-0.4ex}{\small 6})$ and
Eq.~(\ref{eq:stiffgensimp30}), 
to get Eq.~(\ref{eq:highsym4}), as desired.  

\section*{APPENDIX III: REDERIVATION OF EQ.~(\ref{eq:stiffness}) FROM EXACT SOLUTION}
In this appendix, we re-derive Eq.~(\ref{eq:stiffness}) directly from the exact solution for
$\beta(\theta)$ 
given in Eqs.~(\ref{eq:psithetzia}) and (\ref{eq:psiTzia}).  To do so, we just assume
$\cosh\psi_2 \gg \eta_-/\eta_0$
(remember that $\eta_-/\eta_0$ decreases from $\infty$ at $\theta = 0^\circ$ to 1 at $\theta =
30^\circ$, so that, 
between these angles, this condition also implies that $\cosh \psi_2 \gg 1$).  
In this case, Eq.~(\ref{eq:psithetzia}) can be solved to give
\begin{eqnarray}
  \label{eq:psisol}
  \cosh\psi_1 \approx \frac{\eta_0}{\eta_-} \cosh\psi_2. 
\end{eqnarray}
Thus, if $\cosh\psi_2 \gg \eta_-/\eta_0 > 1$, then $\cosh\psi_1 \gg 1$.
We show here that these 
assumptions for $\cosh \psi_{1,2}$, together with the low-temperature 
replacement of $f(z)$ by $1/(2 z)$ in Eq.~(\ref{eq:psiTzia}), 
are enough to derive Eq.~(\ref{eq:stiffness}).
 
When $\cosh \psi_{1,2} \gg 1$, then $\cosh\psi_{1,2} \approx \sinh\psi_{1,2} \approx
e^{\psi_{1,2}}/2$.  With these approximations, Eqs.~(\ref{eq:psithetzia}) and (\ref{eq:psiTzia})
become remarkably simple:
\begin{eqnarray}
  e^{\psi_1} + e^{\psi_2} &=& 2 f(z), \label{eq:psitzialowt} \\
  e^{\psi_2} &=& \frac{\eta_-}{\eta_0} e^{\psi_1} \label{eq:psithetzialowt} .
\end{eqnarray}
Solving this pair of equations for $e^{\psi_1}$ and $e^{\psi_2}$ gives
\begin{eqnarray}
e^{\psi_1} = \frac{2 f(z) \eta_0}{\eta_0+\eta_-}, \qquad
e^{\psi_2} = \frac{2 f(z) \eta_-}{\eta_0+\eta_-}.   \label{eq:psisollowt}
\end{eqnarray}
If we then replace $f(z)$ by its low-temperature limit, $1/(2z)$, Eq.~(\ref{eq:betazia}) becomes
\begin{eqnarray}
  \label{eq:betazialowt}
  \frac{\beta a_{||}}{k_B T} = \eta_0 \ln\left[ \frac{\eta_0}{z(\eta_0+\eta_-)} \right]
                  + \eta_- \ln \left[  \frac{\eta_-}{z(\eta_0+\eta_-)}  \right].\\ \nonumber
\end{eqnarray}

\noindent By noting $\eta_0 + \eta_- = \eta_+$, and using the definition for $z$, 
Eq.~(\ref{eq:betazialowt}) can be easily simplified to Eq.~(\ref{eq:lineTensionTheta}), from 
which Eq.~(\ref{eq:stiffness}) for $\tilde{\beta}$ was derived.

By deriving the approximation given in Eq.~(\ref{eq:lineTensionTheta}) (and thus
Eq.~(\ref{eq:stiffness})) 
in this way, we can determine when the approximation becomes invalid.  
Specifically, we require $\cosh \psi_{1,2} \gg 1$.  As we showed, the 
more restrictive of these inequalities is the one involving $\cosh \psi_1$, since $\cosh \psi_1$
 is necessarily smaller than $\cosh \psi_2$ in the first sextant 
by a factor of $\eta_0/\eta_-$ (which is less than 1).  Thus, the main assumption is   
$\cosh \psi_1 \gg 1$, which, from Eq.~(\ref{eq:psisollowt}), is just 
\begin{eqnarray}
  \label{eq:restrictsol}
  \frac{2 f(z) \eta_0}{\eta_0+\eta_-} \gg 1.
\end{eqnarray}
The solution to this equation, which we call $\theta_2$, is given by the following inequality:
\begin{eqnarray}
  \label{eq:restrictang}
  \cot \theta_2 \ll \frac{4 f -1}{\sqrt{3}}.
\end{eqnarray}
Because $\cot \theta$ decreases from $\infty$ at $\theta = 0$ to $1/\sqrt{3}$ at $\theta =
\pi/6$, we know that 
angles in the first sextant that are greater than 
$\theta_2$ will also satisfy the inequality in Eq.~(\ref{eq:restrictang}).  Thus,
Eqs.~(\ref{eq:lineTensionTheta}) 
and (\ref{eq:stiffness}) 
are valid in the first sextant at all angles above $\theta_2$.

\begin{acknowledgments}
Work at the University of Maryland was supported by the NSF-MRSEC, Grant DMR 00-80008.  TLE
acknowledges partial support of collaboration with ISG at 
FZ-J\"ulich via a Humboldt U.S. Senior Scientist Award.  We are very grateful to R. K. P. Zia
for many insightful and helpful comments and communications.
We have also benefited from ongoing interactions with E.~D.\ Williams and her group.  
\end{acknowledgments}





\begin{thebibliography}{10}
\bibitem{thetanote} The calculation can be facilitated by writing $\eta_i \equiv a_i \cos \theta
+ b_i \sin \theta$ and then using Eq.~(\ref{eq:dt_i}) to show
 $\tilde{s}_i = [(b_i^2-a_i^2 ) ( \cos^2 \theta -\sin^2 \theta) 
- 4 a_i b_i \sin \theta \cos \theta]/[a_i \cos \theta + b_i \sin \theta]$ .
\bibitem{zianote} Eq.~(\ref{eq:psiTzia}) bypasses several intermediate quantities that are
important for Zia's [\onlinecite{zia}] general treatment but cumbersome here.  Our $\psi_2$ has
the opposite sign from Zia's, and our angle $\theta$---which vanishes for close-packed step
orientations---differs by 30$^\circ$ from his.  Note also that interchanging $\psi_1$ and
$\psi_2$  on the LHS of Eq.~(\ref{eq:psithetzia}) inverts the LHS, from which it is easy to see
that $\beta(\theta)$ of Eq.~(\ref{eq:betazia}) is symmetric about $\theta = 30^\circ$. The LHS
of Eq.~(\ref{eq:psiTzia}) is obviously invariant under the interchange of $\psi_1$ and $\psi_2$.
\bibitem{AAnote} In Eqs. (4.30)-(4.33) of Ref.~\onlinecite{AA99}, their $W$ corresponds to our
$z^2$ and their angles, like Zia's [\onlinecite{zia,zianote}], differ from ours so that their
$\pi/2$ is our 0 and their 0 is our $\pi/6$.
\bibitem{tjsnote1} There appears to be a pair of sign or letter typos in Eq. 11 of Ref.
\onlinecite{rottman}.
\end{thebibliography}
\end{document}